# Anomalous Spontaneous Reversal in Magnetic Heterostructures

Zhi-Pan Li[1], Johannes Eisenmenger[2], Casey W. Miller[1] and Ivan K. Schuller[1]

[1] *Physics Department, University of California-San Diego, La Jolla, CA 92093-0319*

[2] *Abteilung Festkörperphysik, Universität Ulm, D-89069 Ulm, Germany*

**Abstract** We observe a thermally induced spontaneous magnetization reversal of epitaxial ferromagnet/antiferromagnet heterostructures under a constant applied magnetic field. Unlike any other magnetic system, the magnetization spontaneously reverses, aligning anti-parallel to an applied field with decreasing temperature. We show that this unusual phenomenon is caused by the interfacial antiferromagnetic coupling overcoming the Zeeman energy of the ferromagnet. A significant temperature hysteresis exists, whose height and width can be tuned by the field applied during thermal cycling. The hysteresis originates from the intrinsic magnetic anisotropy in the system. The observation of this phenomenon leads to open questions in the general understanding of magnetic heterostructures. Moreover, this shows that in general heterogeneous nanostructured materials may exhibit unexpected phenomena absent in the bulk.

Nanoscience has become an active area of research due to the breakdown of expectations when one or more length scales are reduced to the nanoscale. Moreover, nanostructuring combined with proximity effects may produce emergent phenomena that are neither generally found in homogeneous materials nor predicted by simple finite size scaling laws. Conventional semiconductor heterostructures at the nanoscale, result in 2-dimensional electron gases and quantum dots, which exhibit phenomena such as coloumb blockade and the fractional quantum Hall effect [1]. Nanoscale heterostructures of ferromagnets (FMs) with semiconductors, normal metals, and antiferromagnets (AFs) give rise to ferromagnetic semiconductors [2], giant magnetoresistance [3], and exchange bias (EB) [4], which are the basis of the novel field of spintronics [5, 6].

In this Letter, we present a novel and unusual phenomenon in which, under a constant magnetic field, a nanoscale FM in intimate contact with an AF, spontaneously reverses its magnetization with decreasing temperature. This is contrary to the general understanding that an applied field and an electric current are the only two ways to fully reverse the orientation of a ferromagnet's magnetization [7]. We observe that below an upper limit, larger applied magnetic fields induce larger magnetization reversal. Interestingly, the temperature dependence of the magnetization is hysteretic, thus allowing FM switching by thermal cycling.

We observe this phenomenon in exchange bias heterostructures. Exchange bias arises when an AF/FM heterostructure is cooled below the AF Néel temperature, $T_N$, in an external cooling field, $H_{FC}$ [8]. The interaction between the FM and AF across the interface produces a low temperature ($T < T_N$) shift of the magnetization hysteresis (M-H) loop along the magnetic field axis. The shift, or exchange bias field, $H_{EB}$, can be either positive or negative depending on the magnitude of the $H_{FC}$ [9].

Epitaxial exchange biased FM/AF samples were grown on $MgF_2$ (110) substrates by e-beam evaporation with a structure $ZnF_2$ (30nm) / $FeF_2$ (50nm) / FM (3nm) /Al (3nm), with FM = Ni or Co. The $ZnF_2$ is a paramagnetic buffer layer for the epitaxial growth of antiferromagnetic $FeF_2$ ($T_N$ = 78 K). The Al capping layer was used to prevent oxidation. $ZnF_2$ and $FeF_2$ were grown at 300°C, the FM and Al at 150°C, all at 0.05 nm/s with a base pressure of $10^{-7}$ Torr. X-ray diffraction revealed that the $FeF_2$ grows epitaxially untwined in the (110) orientation, while the FM is polycrystalline. The magnetization was measured using superconducting quantum interference device (SQUID) magnetometry with the magnetic field applied parallel to the [001] easy axis of $FeF_2$ in the sample plane. The easy axis of the FM coincides with $FeF_2$ [001]. At $T$ = 10 K, the $FeF_2$ / Ni sample exhibits positive EB with $\mu_0 H_{EB}$ = 0.41 T when cooled in $\mu_0 H_{FC}$ > 0.05 T, and shows both positive and negative EB of $\mu_0 H_{EB}$ = ± 0.41 T for $\mu_0 H_{FC}$ between 0 and 0.05 T (Fig. 1 inset). The $FeF_2$ / Co sample shows solely positive or negative EB for $\mu_0 H_{FC}$ above 0.1 T and below 0.01 T, respectively, with coexistence of both between these fields. Spontaneous reversal only occurs in samples displaying positive EB, either entirely or partially.

At $T$ = 150 K, the $FeF_2$ / Ni sample shows a typical square magnetic hysteresis (M-H) loop [10]. The sample was first saturated at this temperature by a 0.5 T magnetic field, then subject to a constant $H_{FC}$ while saturated. The sample was then cooled to 10 K, then heated to above 150 K. Figure 1 shows the magnetization as the temperature was changed for two constant $H_{FC}$ (M-T curves). For $\mu_0 H_{FC}$ = 0.1 T, the magnetization starts to reverse at 65 K (below $T_N$), reaches zero at 57 K, then fully reverses, aligning anti-parallel to $H_{FC}$ at ~50 K. With increasing temperature, the FM magnetization increases to zero at 104 K, then restores its full alignment with $H_{FC}$ at ~120 K. The net result is a significant M-T hysteresis with a full thermal width at half the reversed magnetization $\Delta T_C$ = 47 K. The hysteresis is more pronounced if cooled in a lower field of $\mu_0 H_{FC}$ = 0.01 T. With this cooling field, the Ni magnetization reverses by 55% relative to full reversal at 63 K, and returns to its original state at 185 K, giving $\Delta T_C$ = 123 K. A similar effect was also observed in the $FeF_2$ / Co sample. At $\mu_0 H_{FC}$ = 0.03 T, the Co magnetization reverses by 68 % at 55 K and switches back at 114 K, giving $\Delta T_C$ = 59 K. This thermal hysteresis is reminiscent of a FM switching between two saturated states in response to a sweeping external magnetic field or electric current [11] at constant temperature.

The change in the magnetization, $\Delta M = M (T = 150$ K$) – M (T = 10$ K$)$, and the width $\Delta T_C$ of the M-T hysteresis can be tuned by $H_{FC}$, as shown in Fig 2. For both FMs, $\Delta M$ increases with increasing $H_{FC}$ until $H_{FC}$ ~0.1 T, after which it decreases until the spontaneous reversal is no longer observed. $\Delta T_C$ rapidly decreases with increasing $H_{FC}$ initially (below ~0.1 T), then slowly tends toward zero for higher $H_{FC}$ (Fig. 2b).

The thermally induced FM reversal results from two competitions: one between the antiferromagnetic (AF) interfacial coupling $\mathcal{H}_{int}$ and the AF Zeeman energy $\mathcal{H}_{AF\text{-}Zeeman}$, and the other between the coupling $\mathcal{H}_{int}$ and the FM Zeeman energy $\mathcal{H}_{FM\text{-}Zeeman}$. The former determines the orientation of the frozen interfacial AF uncompensated moment, $\mathbf{S}_{AF}$, and establishes positive EB; with the AF thus frozen, the latter determines the orientation of the FM.

Positive EB arises when the interfacial AF moment freezes in the magnetic field direction under a cooling field large enough to overcome the AF interfacial coupling [9]. When the interfacial coupling dominates AF Zeeman energy for small cooling fields, the uncompensated AF moment orients opposite to the field and gives rise to negative EB. At intermediate cooling fields, positive and negative EB coexist due to spatially inhomogeneous interfacial coupling. In this case, double hysteresis loops are observed if the length scale of this inhomogeneity is much larger than the FM domain wall width [10, 12]. In our system, 50% of the sample exhibits positive EB for $H_{FC}$ = 0.005 T. The origin of this surprisingly low onset cooling field for positive EB is thus far unknown.

Once positive EB is established, the reversal of the FM is governed by the competition of the FM Zeeman energy with the AF interfacial coupling. The FM Zeeman energy favors the FM aligning parallel to $H_{FC}$, while positive frozen $S_{AF}$ and the AF interfacial coupling favors an antiparallel orientation. $S_{AF}$, and thus $\mathcal{H}_{int}$, increases as the AF becomes increasingly ordered with decreasing temperature below $T_N$, as evidenced by the increase of $H_{EB}$ (Fig. 3 inset). $\mathcal{H}_{int}$ eventually overcomes the FM Zeeman energy, causing the FM to spontaneously reverse its magnetization. In the case of purely negative EB (negative $S_{AF}$), the AF interfacial coupling assists in aligning the FM magnetization parallel to $H_{FC}$, and thus will not lead to spontaneous reversal.

This competition can also explain the unusual low field behavior: i.e. increasing $H_{FC}$ causing $\Delta M$ to increase (Fig. 2). When a larger $H_{FC}$ is applied, the positively EB regions of the sample increase in area at the cost of negatively EB regions. As a result, a larger percentage of the FM reverses. If the field is large enough that the entire sample exhibits positive EB, $S_{AF}$ can no longer increase. In this case, with increasing field and thus increasing FM Zeeman energy, $\Delta M$ decreases and ultimately vanishes when $\mathcal{H}_{FM\text{-}Zeeman} > \mathcal{H}_{int}$.

Quantitatively, the two competing energies that govern the reversal process can be expressed as $\mathcal{H}_{FM\text{-}Zeeman} = -\mu_0 H_{FC} M_{FM} t_{FM}$, and $\mathcal{H}_{int} = -J_{FM/AF} S_{AF} \cdot S_{FM}$, where $M_{FM}$ and $t_{FM}$ are the magnetization and thickness of the FM, respectively, $S_{AF}$ and $S_{FM}$ are the AF and FM interfacial moment per unit interface area, and $J_{FM/AF} < 0$ is the interfacial coupling between the AF and FM. In order for the interfacial coupling to reverse the FM magnetization, it also has to overcome an energy barrier, $\mathcal{H}_{barrier}$, between the two saturated states of the FM. This energy barrier is determined by the intrinsic FM anisotropy, anisotropy induced by the interfacial coupling, and the energy related to domain formation. Spontaneous reversal occurs when $|\mathcal{H}_{int}| > |\mathcal{H}_{FM\text{-}Zeeman}| + \mathcal{H}_{barrier}$, and aligns with the field when $|\mathcal{H}_{FM\text{-}Zeeman}| > |\mathcal{H}_{int}| + \mathcal{H}_{barrier}$. Adopting the Meiklejohn-Bean model [4, 8] allows us to rewrite the interfacial coupling as $\mathcal{H}_{int} = \pm \mu_0 H_{EB} M_{FM} t_{FM}$, where the sign refers to the sign of $S_{FM}$. Thus for negligible $\mathcal{H}_{barrier}$, the reversal should occur when $H_{EB}(T) = H_{FC}(T)$ for both cooling and heating, without any hysteresis. Figure 3 shows $H_{EB}$ and $H_{FC}$ as functions of temperature for $FeF_2$ / Ni with $\mu_0 H_{FC} = 0.1$ T. The condition $H_{EB} = H_{FC}$ is satisfied at point C with $T = 70$ K. Experimentally, $\mathcal{H}_{barrier}$ is not negligible evidenced by the significant coercivity $H_C$ enhancement around $T_N$ (Fig. 3). This $H_C$ enhancement is attributed to short-range order in the AF [13, 14]. Using $\mathcal{H}_{barrier} = \mu_0 H_C M_{FM} t_{FM}$, the reversal condition becomes $H_{EB}(T) \sim H_{FC} \pm H_C(T)$, where positive and negative signs refer to cooling and heating, respectively. This leads to a lower reversal temperature for cooling and higher for heating than predicted by $H_{EB} = H_{FC}$. Figure 3 shows that the reversal condition is satisfied at 57 K (point A) and 105 K (point B), for cooling and heating, respectively, in agreement with Fig. 1. While the details of the reversal process are unknown, this shows that the M-T hysteresis with a tunable width $\Delta T_C$ originates from the temperature dependent interface-induced anisotropy.

The interfacial coupling energy must dominate the FM Zeeman energy for FM spontaneous reversal. This condition is experimentally realized using FMs with nanoscale thickness because $\mathcal{H}_{FM\text{-}Zeeman}$ (proportional to $t_{FM}$) can be tuned to be on the order of $\mathcal{H}_{int}$, which is thickness independent. Thus, increasing the FM thickness should lead to lower spontaneous reversal temperatures until the phenomenon disappears. In this case, the FM magnetization can no longer fully reverse; it may still exhibit reversal tendencies such as spontaneous rotation or domain formation with decreasing temperature.

To investigate this further, vector SQUID magnetometry was used to measure the longitudinal (parallel to $H_{FC}$) and transverse (perpendicular to $H_{FC}$ in the sample plane) components of the magnetic moment of a sample with 21 nm thick Ni on $FeF_2$. The temperature dependence of $H_C$ shows that this sample exhibits a low reversal energy barrier: the peak coercivity $\mu_0 H_C = 0.015$ T at $T = 90$ K was small compared to 0.18 T for the 3 nm thick Ni samples. The approximate reversal condition $H_{EB} \sim H_{FC}$ is thus appropriate here. The two components were measured while cooling from $T = 150$ K to 10 K in $\mu_0 H_{FC} = 0.2$ T, and heating in the same field. In this cooling field, the sample exhibits positive EB with $H_{EB} = 0.1$ T. Coexistence of positive and negative EB is encountered for $\mu_0 H_{FC}$ between 0.1 and 0.2 T, while only negative EB exists for $\mu_0 H_{FC}$ less than 0.1 T. Since $H_{EB} < H_{FC}$ for $\mu_0 H_{FC} = 0.2$ T, the interfacial coupling cannot overcome the FM Zeeman energy, and thus no spontaneous reversal should be observed. The measurement showed a small reduction of the longitudinal and a large increase of transverse moment with decreasing temperature, with the total magnetic moment above 0.96 $M_S$ (Fig. 4). Therefore, although unable to fully reverse as in thin FMs, here the FM magnetization nearly coherently rotated away from the magnetic field direction by about 30 degrees. This FM spontaneous rotation was not hysteretic due to the small $H_C$, signature of small intrinsic and AF-induced anisotropy of the FM. A larger cooling field reduces the amount of rotation, similar to the behavior in the high field range of Fig. 2.

Although the above discussion explains the observed phenomenon, and correctly gives an estimate of reversal temperatures, it also leads to important open questions. The present understanding of positive EB implies that $|H_{AF-Zeeman}| > |H_{int}|$ below $T_N$. At the same time, the FM reversal condition requires $|H_{FM\text{-}Zeeman}| < |H_{int}|$ below the FM reversal temperature. Therefore, $|H_{AF\text{-}Zeeman}| > |H_{FM\text{-}Zeeman}|$ below the reversal temperature, or $m_{AF} > m_{FM}$, where $m_{AF}$ refers to the uncompensated frozen AF moment. It is reasonable that when $m_{AF}$ becomes larger than $m_{FM}$, the FM should reverse with $m_{AF}$ in the field direction. This is similar to some ferrimagnet Gd-Co [15] and multilayer systems Co/Gd [16], which results from two antiferromagnetically coupled spin species competing to align with the field. In this case, the two magnetizations can be clearly identified and its total moment at a low enough temperature is always positive. However, in our FM/AF system, a large $m_{AF}$, would manifest as a significant shift of the M-H loop [17] along the magnetization axis, which was not observed (Fig. 1 inset). This suggests that $m_{AF}$ is much smaller than $m_{FM}$, contrary to the previous argument, yet spontaneous reversal still occurs with a *negative* low-temperature magnetic moment for certain cooling fields.

In summary, we report a novel temperature-driven phenomenon where, under a constant applied magnetic field, saturated magnetic heterostructures spontaneously reverse their magnetization. This phenomenon is observed when the heterostructure exhibits positive exchange bias. This reversal behavior shows a significant temperature hysteresis that can be tuned by the field applied during thermal cycling, in contrast to the conventional temperature-dependent hysteretic behavior of a FM under magnetic field cycling. This behavior not only provides another means for inducing ferromagnetic reversal beside magnetic fields and electric current, but also offers possible probes for buried interfaces and AF. Although the proposed interpretation is able to partially explain the phenomenon, it also leads to open questions due to our incomplete understanding of exchange bias in general.

Work supported by US-DOE. Financial support of Cal-(IT)$^2$ (Z.- P. L.) is acknowledged.

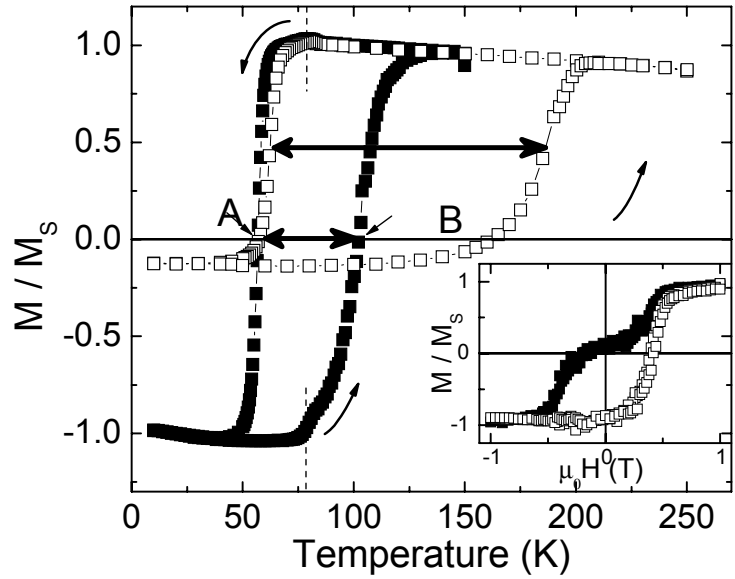

Fig. 1 Normalized magnetization of FeF$_2$ (50 nm) / Ni (3 nm) measured under temperature sweep in 0.1 T (solid squares) and 0.01 T (empty squares) by SQUID magnetometry. The dashed line marks $T_N$ = 78 K of FeF$_2$. The width $\Delta T_C$ is marked by thick horizontal arrows at <$M$>. Points A and B are the reversal temperatures 57 K and 104 K for $\mu_0 H_{FC}$ = 0.1 T (see Fig. 3 for more details). (Inset) Magnetization hysteresis loops for FeF$_2$/Ni at T = 10 K for $\mu_0 H_{FC}$ = 0.01 T (solid squares) and 0.1 T (empty squares).

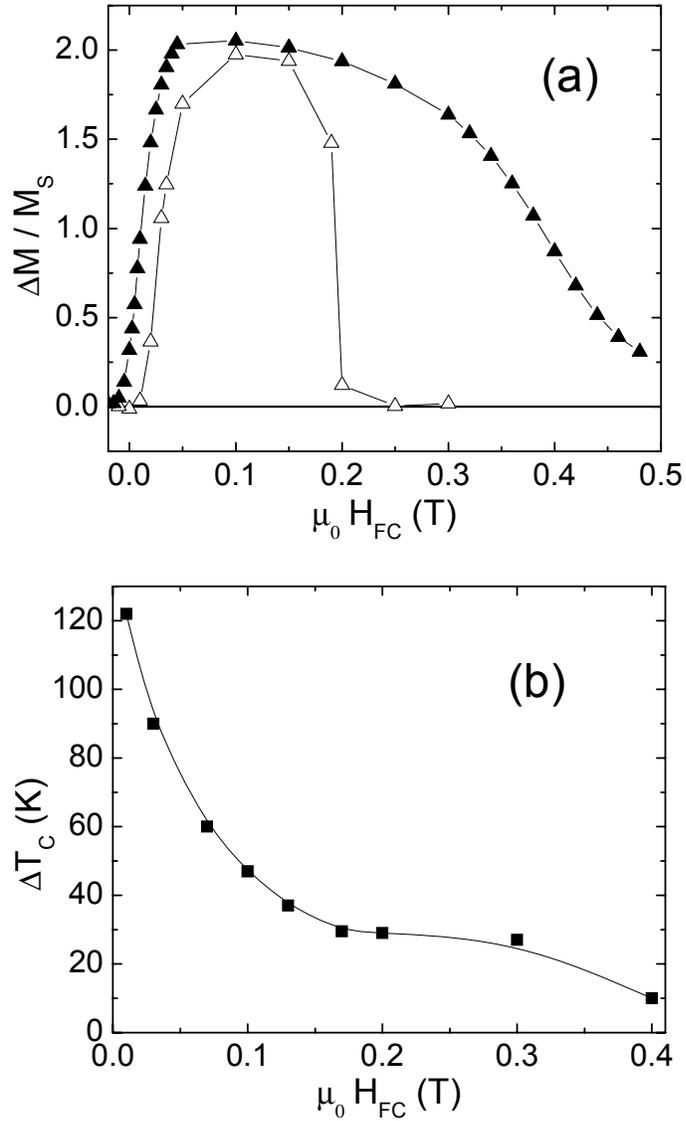

Fig. 2 (a) Magnetic cooling field dependence of the magnetization change $\Delta M$ during fast thermal cycling normalized by the saturation magnetization $M_S$ (insert a value) for $FeF_2$ / Ni (solid triangles) and $FeF_2$ / Co (open triangles). A cooling speed dependence results in systematic and controllable differences of up to 10% in $\Delta M$. (b) Magnetic field dependence of the full thermal width $\Delta T_C$ at $<M>$. Lines are guides to the eye.

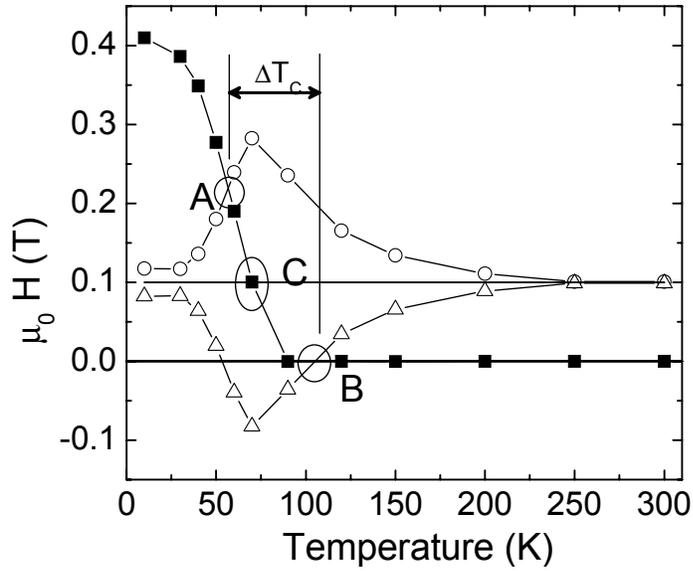

Fig. 3. Exchange bias $H_{EB}$ (solid squares), $H_{FC} + H_C$ (empty circles), $H_{FC} - H_C$ (empty triangles) as functions of temperature. The cooling field $H_{FC}$ is marked by the horizontal line at 0.1 T. Points A and B mark the reversal temperatures with FM anisotropy considered, in agreement with the position of points A and B in figure 1. Point C refers to the reversal temperature for a negligible FM reversal barrier.

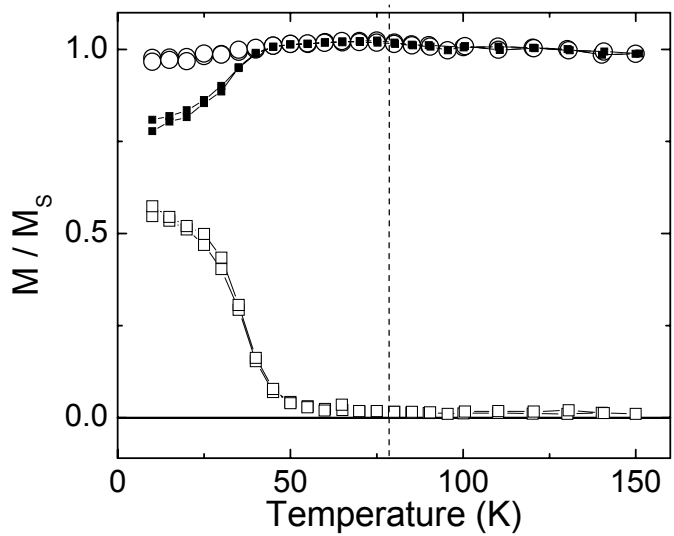

Fig. 4. Normalized in plane longitudinal (solid squares), transverse (empty squares) and total (empty circles) magnetization of $FeF_2$ (50 nm) / Ni (21 nm) measured by vector SQUID magnetometry in thermal cycling with a 0.2 T magnetic field. $T_N$ is marked by the dashed line.